\documentclass[11pt]{article}

\usepackage{graphicx}
\usepackage{multicol}
\usepackage{amsmath,amssymb,cite,color,hyperref}
\newcommand{\be}{\begin{equation}}
\newcommand{\ee}{\end{equation}}
\newcommand{\bea}{\begin{eqnarray}}
\newcommand{\eea}{\end{eqnarray}}
\newcommand{\bi}{\begin{itemize}}
\newcommand{\ei}{\end{itemize}}
\newcommand{\ben}{\begin{enumerate}}
\newcommand{\een}{\end{enumerate}}

\def\frac#1#2{{{#1}\over {#2}}}
\def\gsim{\mathrel{\rlap{\lower4pt\hbox{\hskip1pt$\sim$}}
    \raise1pt\hbox{$>$}}}         
\def\lsim{\mathrel{\rlap{\lower4pt\hbox{\hskip1pt$\sim$}}
    \raise1pt\hbox{$<$}}}         

\newcommand{\draft}[1]{}

\definecolor{grey}{rgb}{0.5,0.5,0.5}

\usepackage{cite}
\usepackage{hyperref}
\usepackage{url}

\begin{document}

\begin{center}
{\large\bf PDF uncertainties in the extraction of $M_W$ at the LHC:\\
a Snowmass Whitepaper}
\vspace{0.6cm}

Juan~Rojo$^1$  and Alessandro Vicini $^2$

\vspace{.3cm}
{\it ~$^1$ PH Department, TH Unit, CERN, CH-1211 Geneva 23, Switzerland \\
~$^2$ Dipartimento di Fisica, Universit\`a di Milano and
INFN, Sezione di Milano,\\ Via Celoria 16, I-20133 Milano, Italy}
\end{center}   

\vspace{0.2cm}

\begin{center}
{\bf \large Abstract}
\end{center}

The precision measurement of the 
$W$ boson mass is an important milestone for the LHC physics program
in the coming years.
An accurate measurement of $M_W$ allows to
perform stringent consistency tests of the Standard Model by means
of global electroweak fits, which in turn are sensitive to New Physics at
scales potentially higher than the ones explored in direct searches.
From the theoretical point of view, our limited knowledge of PDFs
 will be one of the
dominant sources of uncertainty in ongoing and future LHC
determinations of $M_W$.
In this whitepaper, we have quantified the impact
of PDF uncertainties in the  $W$ mass extractions from the
transverse mass distribution at the LHC.
The calculation has been performed using the NNPDF2.3 set,
which includes direct constrains on the $W$ boson
production kinematics with data for electroweak gauge boson production from
the LHC.
Our results confirm previous estimates that PDF uncertainties 
in the determination of $M_W$ from the $m_W^T$ distribution
are moderate, around 10 MeV at most.
We briefly discuss also the case of the lepton $p_T$ distribution.

\clearpage

The very precise measurement of the $W$ mass at the Tevatron~\cite{Group:2012gb} allows to
perform stringent consistency tests of the Standard Model by means
of global electroweak fits, which in turn are sensitive to New Physics at
scales potentially higher than the ones explored in direct searches.
The continuous improvement in 
experimental accuracy for the extractions of $M_W$ at hadron colliders
 implies that 
parton distribution functions (see~\cite{Ball:2012wy} for a recent benchmark comparison) are  now one of the dominant 
theoretical uncertainties.
With this motivation, it becomes important  to quantify in detail
 the role of PDF uncertainties  and
to explore new avenues to reduce their impact in the final 
$M_W$ determination.

In order to determine the contribution to $\delta M_W$ arising from PDF uncertainties at hadron colliders, Ref.~\cite{Bozzi:2011ww} performed an analysis using a template-fit
method, 
a strategy similar to the one used by the Tevatron collaborations.
 In a first step, we generated pseudo-data for the
$W$ transverse mass distribution $m_T^W$, for various NLO PDF sets and their corresponding
error sets: CTEQ6.6~\cite{Nadolsky:2008zw}, MSTW08~\cite{Martin:2009iq} 
and NNPDF2.1~\cite{Ball:2011mu}. 
Then, with the central
CTEQ6.6 as input, we generated a large number of accurate templates varying
$M_W$ from the reference value in a suitable range. 
The shift in $M_W$ corresponding to each PDF error set is determined
by the template which leads to a better $\chi^2$ agreement with the
corresponding pseudo-data.
Templates were generated with {\tt HORACE}~\cite{CarloniCalame:2006zq,CarloniCalame:2007cd} 
at LO and with
 {\tt DYNNLO}~\cite{Catani:2010en} at NLO.
 
\begin{figure}[h]
\centering
\includegraphics[width=0.62\hsize]{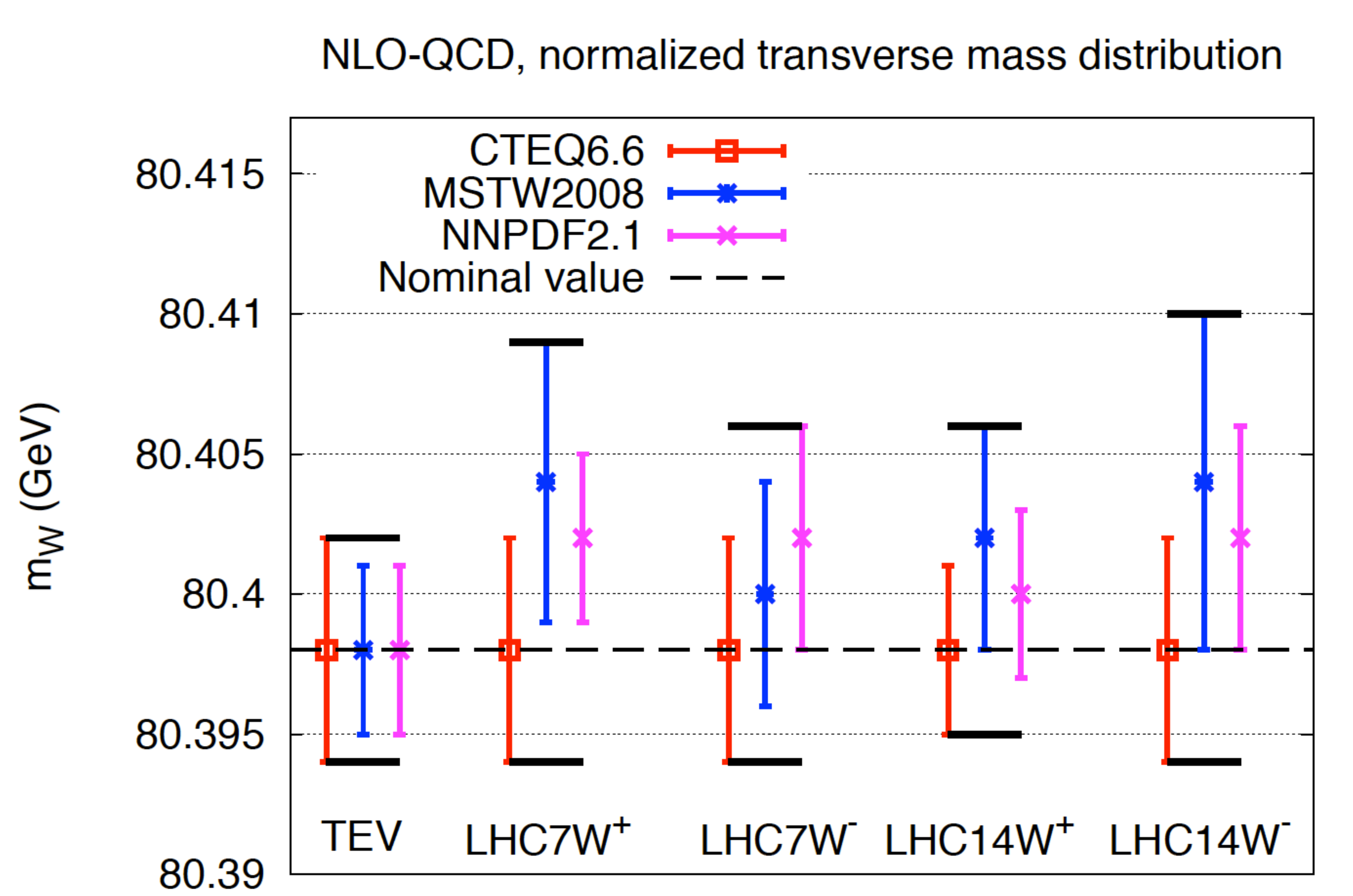}
\caption{\small  PDF uncertainties in the $W$ boson
determination from the $m_T^W$ distribution at the
Tevatron and the LHC, taken from Ref.~\cite{Bozzi:2011ww}. The templates for different $M_W$ values have been
generated with the CTEQ6.6 set.
}
\label{fig:wmass}
\end{figure}

An essential ingredient of our approach was to normalize the templates
to the total integral of the distribution in the fit
region, since in this way the PDF uncertainty is substantially reduced, 
without losing sensitivity to the value of $M_W$.
In fact, the PDF sensitivity enters
mostly through the normalization of the $m_T^W$ distribution, while the sensitivity
to $M_W$ arises from the shape, and is quite independent of the overall normalization.
Our results from  Ref.~\cite{Bozzi:2011ww} are summarized in Fig.~\ref{fig:wmass}. 
The conservative estimate was that PDF uncertainties in $M_W$
fits at the LHC would not be larger than 20 MeV.
Our analysis did not support previous studies~\cite{Krasny:2010vd}, 
which claimed
that achieving a 10 MeV accuracy on $M_W$ at the LHC was
out of reach precisely due to PDF uncertainties.

None of the PDF sets studied in Ref.~\cite{Bozzi:2011ww}
included the recent constraints from the LHC measurements of electroweak
boson production, which determine the PDFs for the same flavor combinations and
the same kinematical regions relevant for $M_W$ determinations.
In order to update our study taking this information into account,
we have revisited the determination of $\delta M_W$ at NLO-QCD of
 Ref.~\cite{Bozzi:2011ww},
with    {\tt DYNNLO} for the theory modeling,
but now 
the NNPDF2.3 set~\cite{Ball:2012cx}.
 This set is particularly suited for the determination
of the $W$ mass at the LHC since it already includes constraints
from $W$ and $Z/\gamma^*$ production data from ATLAS, CMS and LHCb.

In order to reduce the statistical fluctuations,
a dedicated set of $N_{\rm rep}=1000$ replicas of NNPDF2.3 NLO has been produced
and used to compute the theory predictions for the $W$ transverse mass
distribution at NLO-QCD.
In addition, the templates have been computed with very high-statistics runs,
in order to tame Monte Carlo fluctuations as much as possible,
below the size of PDF uncertainties.
All the results shown below  correspond to
$W^+$ production, but as discussed in~\cite{Bozzi:2011ww}, they should apply
in a similar way to $W^-$ production. 

\begin{figure}[h]
\centering
\includegraphics[scale=0.48]{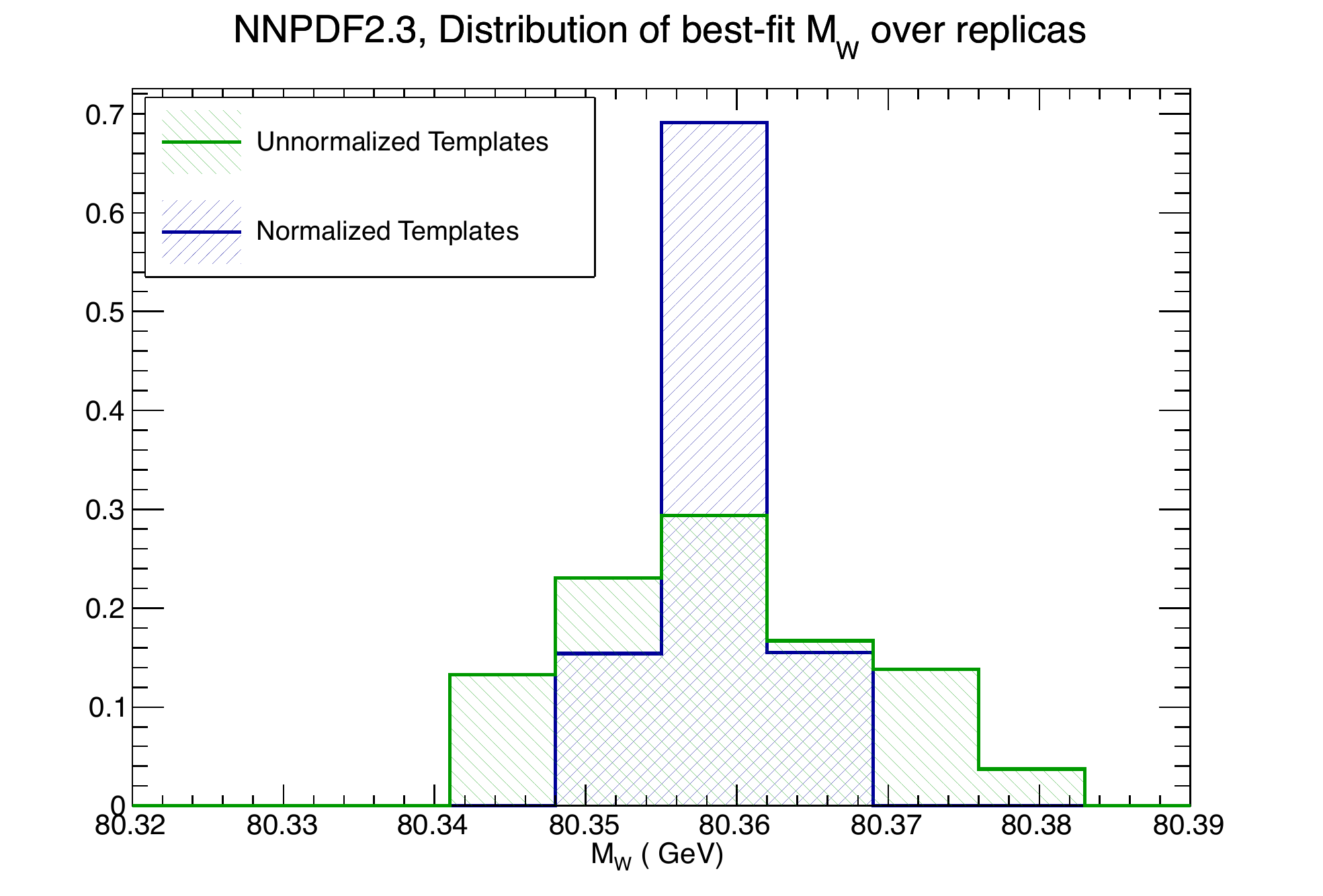}
\caption{\small The distribution of best-fit $M_W$ values
obtained from the comparison of the 1000 replicas of NNPDF2.3
with the reference templates.
We show the results both in the case of unnormalized and
in the case of normalized templates.
 }
\label{fig:histo}
\end{figure}

\begin{table}[h]
\centering
\begin{tabular}{c|c|c}
\hline
&   Absolute Templates  & Normalized Templates  \\
\hline
$M_W\pm \delta_{\rm pdf} M_W$  &  $80.359 \pm 0.010 $ &   $80.359 \pm 0.004 $ \\
\hline
\end{tabular}
\caption{\small \label{tab1} PDF uncertainties in the determination of $M_W$ 
at the LHC 7 TeV using the template fits to the $m_T^W$ distribution 
with NNPDF2.3 $N_{\rm rep}=1000$.
}
\end{table}

For each of the $N_{\rm rep}=1000$ PDF replicas, the fit determines
which is the value of $M_W$ which maximizes the agreement with the 
template distributions. The distribution of best-fit $M_W$ over
the NNPDF2.3 replicas, for both cases of normalized and
unnormalized distributions, is shown in Fig.~\ref{fig:histo}.
It is clear that the distribution obtained from the normalized templates
is narrower, indicating reduced sensitivity to PDF uncertainties.
The mean and the width of these histograms are reported in 
Table~\ref{tab1}: PDF uncertainties are halved when normalized templates are
used, and an uncertainty of $\delta_{\rm pdf}M_W\sim 4$ MeV with NNPDF2.3 is found in this case.
Note also that our estimate applies to the $m_T^W$ distribution fits only,
and to derive the final result we would need as well the results obtained with the MSTW and CT10 sets, which can lead to an increase of the total PDF uncertainty 
by up  to a factor two.

\begin{figure}[h]
\centering
\includegraphics[scale=0.36]{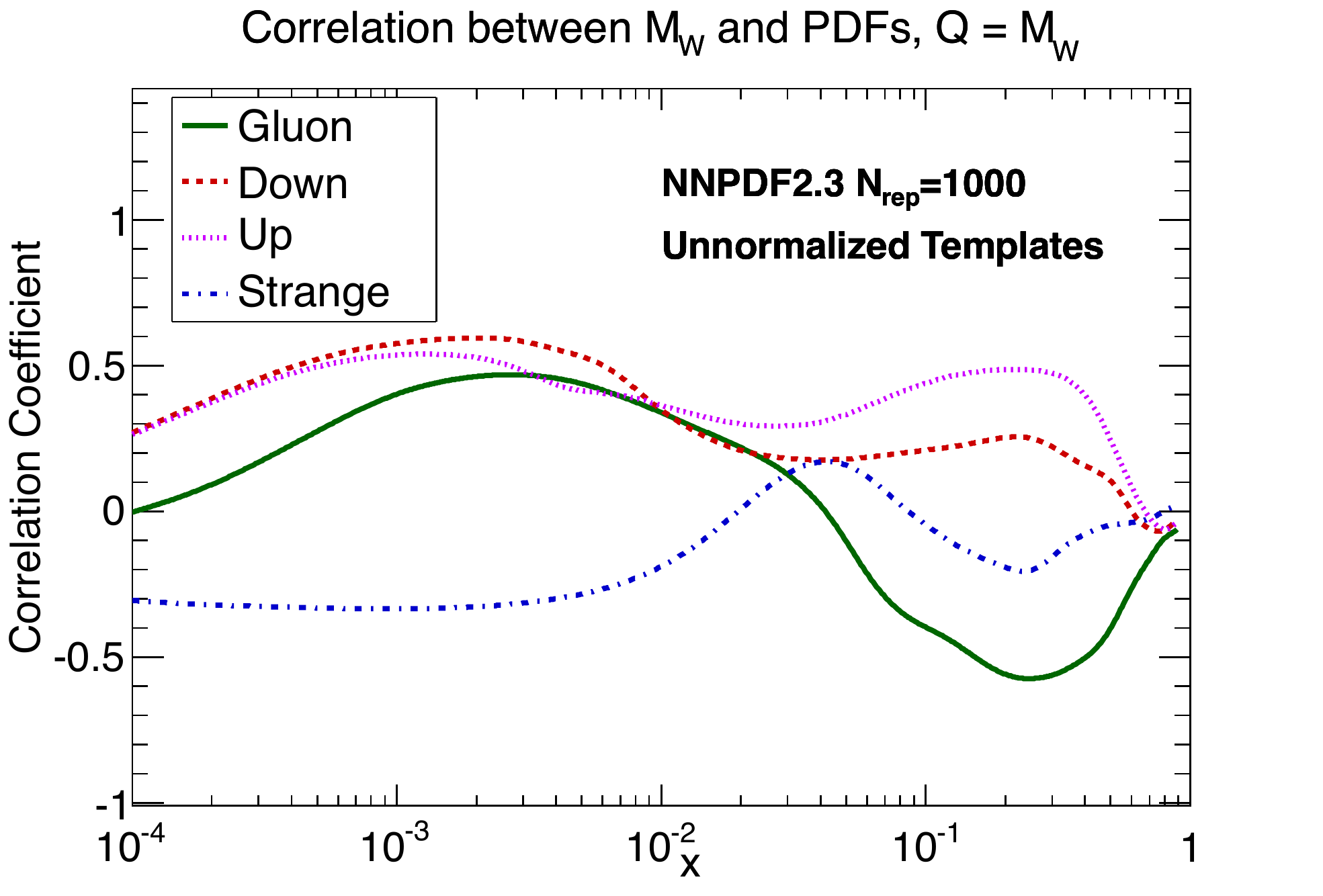}
\includegraphics[scale=0.36]{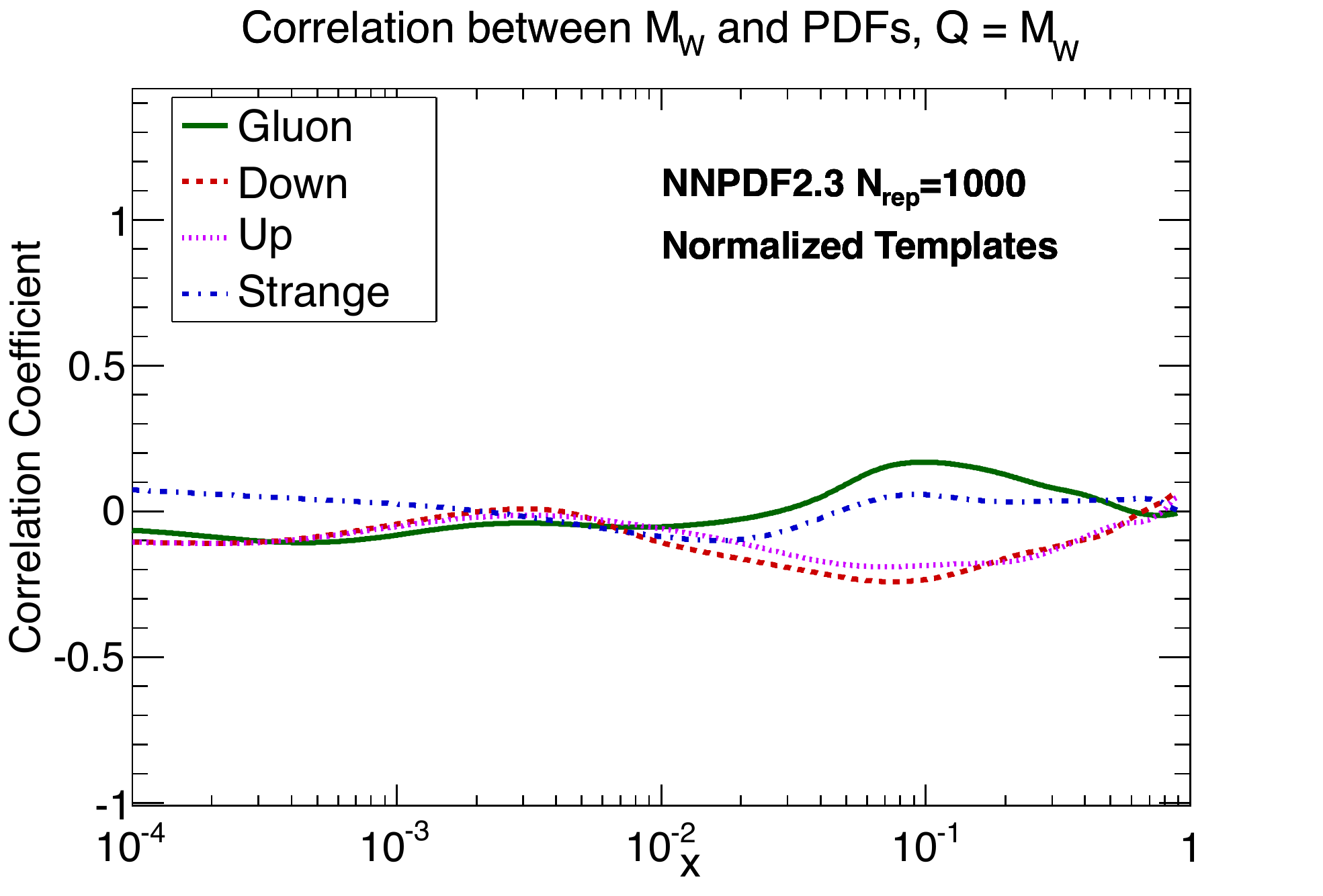}
\caption{\small  Correlations between different PDF flavours
and the $M_W$ determination at LHC 7 TeV, as a function of
Bjorken-$x$, for unnormalized (left plot) and normalized (right plot)
templates.
The predictions from the 1000 replicas of NNPDF2.3 have been used
in the computation.
}
\label{fig:correlations}
\end{figure}

In order to determine if a particular PDF combination is responsible
for the bulk of the PDF uncertainties in $M_W$, it is useful to compute the
correlations~\cite{Demartin:2010er} 
between the $N_{\rm rep}=1000$ PDF replicas of NNPDF2.3 and the 1000
determinations of $M_W$ obtained from the template fits for each replica.
The results are shown in Fig.~\ref{fig:correlations} for the unnormalized
templates (left plot) and for the normalized templates (right plot).
In the case of the unnormalized templates, the correlation between PDFs and $M_W$
is similar to the case of the inclusive $W^+$ cross section~\cite{Ball:2011mu}.
On the other hand, for the normalized templates the correlations are much smaller,
showing that the normalization effectively decorrelates the $M_W$ fits
with respect to the PDFs.
It is clear that
there is not
a particular range of  Bjorken-$x$ or a particular quark flavor
that dominates the $M_W$ measurement. 
This implies that, in order to further reduce the PDF uncertainty in the $M_W$ measurement from $m_T^W$,
one needs new data constraining all quark flavors and gluons in the broadest possible $x$ range.

The  results that we have just discussed were based on the 
determination of $M_W$
from the $W$ transverse mass distribution.
This distribution receives small higher-order QCD corrections,
but its accurate measurement at the LHC will be challenging,
in view of a competitive $M_W$ measurement.
Now we report on
preliminary work towards the extension of the results of~\cite{Bozzi:2011ww}
to a template-fit analysis of the lepton transverse momentum distribution,
which has also been successfully used at the Tevatron to measure $M_W$.

At variance with  the transverse mass distribution, the lepton transverse momentum, $p_T^l$, is
substantially modified by higher-order QCD corrections, given its 
strong correlation with the $W$ boson transverse momentum, $p_T^W$.
 For this distribution the use of resummed calculations for the
$W$ boson $p_T^W$ is required, either using analytical $p_T^W$ QCD resummation
\cite{Bozzi:2010xn}
or NLO-(QCD+EW) calculations matched to (QCD+QED) parton showers
\cite{Barze:2012tt,Barze':2013yca}, 
with a significant increase in the amount of CPU time needed to generate the theory templates.

\begin{figure}[h]
\centering
\includegraphics[scale=0.13]{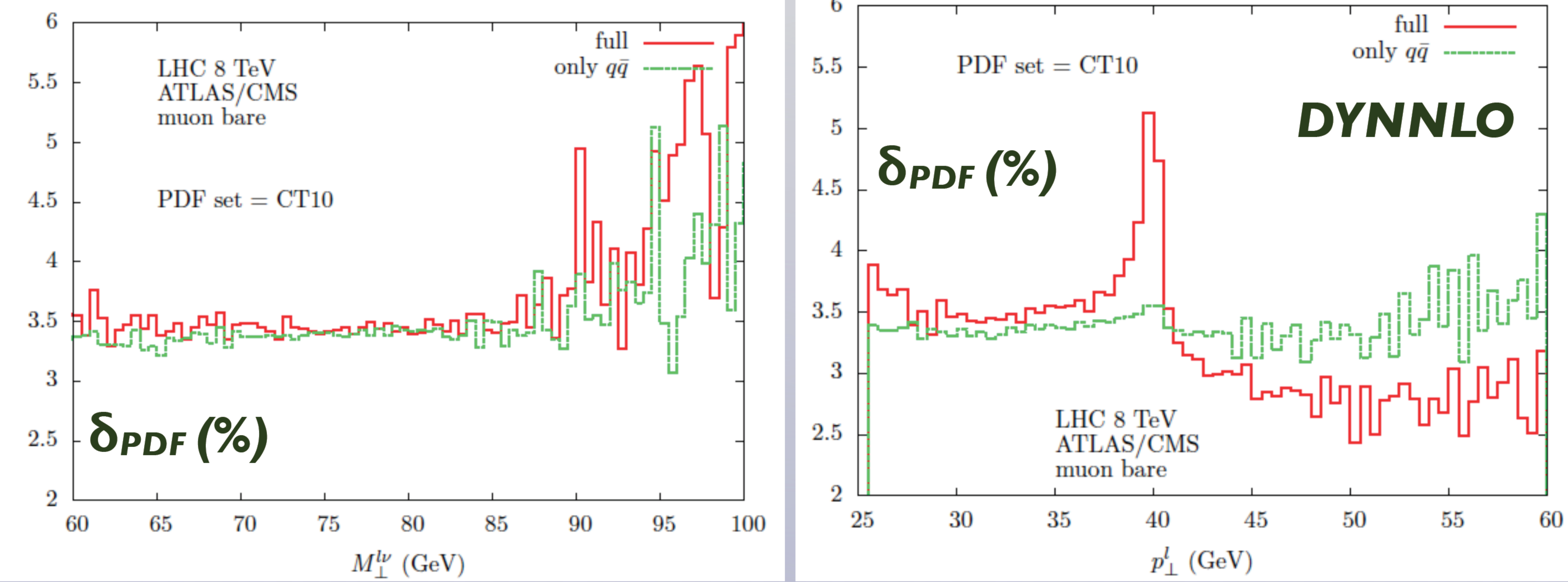}
\caption{\small  The total relative PDF
uncertainty and the separate contribution of quark-antiquark
diagrams  for the transverse
mass (left plot) and the lepton $p_T$ (right plot) distributions,
computed at NLO with {\tt DYNNLO}.
}
\label{fig:pt}
\end{figure}

The relevance of NLO-QCD corrections implies that the gluon PDF
leads also to a more important contribution to the PDF uncertainty on
$M_W$ than in the transverse mass case. 
In order to confirm this,
in Fig.~\ref{fig:pt} we show the contribution of quark-antiquark
terms to the total PDF uncertainty in the transverse mass and lepton
$p_T$ distributions, computed at NLO-QCD with {\tt DYNNLO}.
Therefore, for the lepton $p_T^l$ distribution 
the contribution of the quark-gluon subprocess is substantial, in particular
near the Jacobian peak.

It should be stressed that the results shown in Fig.~\ref{fig:pt} have been obtained at fixed order, whereas, as mentioned above,
 a fully resummed calculation is necessary in the lepton $p_T^l$ case; 
furthermore, the quark-antiquark contribution alone provides only 
a correct estimate of its PDF uncertainty; 
only the results that include all the partonic subprocesses are sensible in terms of physical distributions.
With these two caveats in mind, it is clear that a dedicated analysis should be pursued
in order to limit as much as possible the contribution to
$\Delta M_W$ due to the gluon PDF.
For example, ratios of $W$ over $Z$ distributions
provide a significant cancellation of contributions which are common in the two cases, 
such e.g. the quark-gluon initiated subprocesses,
strongly reducing the corresponding contribution to the PDF uncertainty.

As a final remark, let us mention that PDFs with QED contributions included should be taken
into account to consistently assess the corrections to the $M_W$ fits induced by QED effects.
The recently released NNPDF2.3 QED set~\cite{Ball:2013hta} 
is especially suitable for this purpose, since
it includes  NNLO QCD combined with LO QED corrections, and also the most recent constraints
from electroweak gauge boson production data at the LHC.
The implications of NNPDF2.3 QED for $M_W$ determinations should be the topic of detailed
studies in the near future.

To conclude, 
LHC is collecting a huge amount of high-quality data that should be fully 
exploited in order to improve our knowledge of the proton parton densities.
Taking into account present and future information on PDFs, 
as well as recent improvements in theory modeling,
the  measurement of the $W$ mass at the LHC with a PDF
uncertainty as small as $\delta_{\rm pdf} M_W \sim 5-10$ MeV
is certainly within reach.
The experimental uncertainties are expected to be of a similar size.
To further improve on this result, future lepton colliders will be 
required, such as the recent TLEP proposal~\cite{Gomez-Ceballos:2013mfa}.


\end{document}